\documentstyle[pra,aps,epsf]{revtex}

\begin{document}

\title{Helium nanodroplets and trapped Bose-Einstein condensates \\
as prototypes of finite quantum fluids}

\author{Franco Dalfovo$^1$ and Sandro Stringari$^2$}

\address{$^1$ Dipartimento di Matematica e Fisica, Universit\`a
Cattolica del Sacro Cuore, and \\
Istituto Nazionale per la Fisica della Materia, Unit\`a di Brescia, 
Brescia, Italy }

\address{$^2$ Dipartimento di Fisica, Universit\`a di Trento, and \\
Istituto Nazionale per la Fisica della Materia, Unit\`a di Trento, 
I-38050 Povo, Italy }

\maketitle

\date{ \today }

\begin{abstract}
Helium nanodroplets and trapped Bose-Einstein condensates in
dilute atomic gases offer complementary views of fundamental
aspects of quantum many-body systems. We discuss analogies and
differences, stressing their common theoretical background and
peculiar features. We briefly review some relevant concepts, such
as the meaning of superfluidity in finite systems, the behavior
of elementary excitations and collective modes, as well as 
rotational properties and quantized vorticity.   
\end{abstract}

\section{Introduction}
\label{sec:intro}

Helium clusters have been the object of a rather extensive
investigation in the last two decades. They are becoming even more
interesting nowadays, due to the improved accuracy of recent
experiments where several properties of the clusters are characterized
by using dopant atoms and molecules as efficient probes \cite{jcp2001}.
In 1995, the observation of Bose-Einstein condensation (BEC) in
ultracold vapors of alkali-metal atoms 
\cite{bec95-jila,bec95-mit,bec95-hulet} opened
another active field of research. This event is seemingly
disconnected from the physics of helium clusters. The framework
in which  BEC was initially investigated is a mix of atomic physics
and quantum optics, while helium clusters are a typical subject of
condensed matter and chemical physics. As a consequence, some
of the key words used to interpret the first experimental results
on BEC (like order parameter, coherence length, phase coherence,
elementary excitations, and others) had sometimes different meaning
in the two communities. However, as soon as trapped condensates
were sufficiently characterized, by using the Gross-Pitaevskii
theory as a starting point, most of the physics behind these systems
appeared to be strongly connected with the physics of superfluids.
One can even say that helium droplets and trapped atomic gases
correspond to two limiting cases of the same system since they are,
respectively, examples of very dense and dilute finite-sized quantum
fluids. Despite the different type of confinement (self-binding 
for helium droplets and external trapping for atomic gases) these
systems belong to the same conceptual framework, possibly based on a 
common language.

A quantum fluid is a system of many interacting particles for which
one can locally define both a density and a velocity field, whose
behavior is strongly affected  by
quantum correlations. An example of these quantum effects is the zero
point motion, which is crucial in determining the typical lengthscales
for density modulations, like the surface thickness at the fluid
boundaries, the core size of a vortex, the size of soliton-like
structures. Another important effect is the fact that the
fluid, or part of it, moves with a non-dissipative and irrotational
velocity field, which means that the system is superfluid.
These and other features can be associated with the existence of
an {\sl order parameter}. Qualitatively, as the particles are cooled 
down to enough low temperatures their de Broglie wavelength becomes 
larger than their average separation, so that they loose their
individual identity. In the case of bosons, they can  be treated as 
components of a single ``macroscopic wave function".  The concept of 
order parameter is the appropriate tool to implement this qualitative 
idea.

The  temperature at which these effects occur is generally very low, 
so that the fluid actually freezes before
reaching this regime. In the case of liquid $^4$He, the light atomic
mass allows the system to remain liquid down to zero temperature and
the system becomes a quantum fluid for temperatures lower than about
$2.2$ K. In the case of trapped condensates, vapors of alkali-metal
atoms are kept in a metastable gas-phase down to temperatures of the
order of tens of nanokelvin, where quantum effects becomes dominant.
The fact that these systems are in a metastable configuration
is compatible with the existence of a kinetic equilibrium which
is ensured by two-body elastic collisions. The transition to the
thermodynamic equilibrium, given by the crystal phase, is instead
driven by three-body recombinations, which are however rare events in very
dilute samples and take place on a relatively long time scale.  

When the fluid becomes a quantum fluid the type of quantum
statistics is also crucial and one can find either a Fermi or a Bose
fluid with very different properties. Liquid $^3$He is the fermionic
counterpart of $^4$He, and also alkali-metal atoms with both Fermi and
Bose statistics can be trapped and cooled down to the quantum
regime. In this brief review, however, we concentrate mainly on 
Bose fluids.

The paper is organized as follows. First we remind the basic
notions of Bose-Einstein condensation, as a common starting point.
Then we discuss two opposite limits: a dilute and cold
gas, where one can develop a rigorous theory for interacting bosons,
and a dense quantum fluid, where more sophisticated many-body
techniques are required. Once the theoretical framework is
sketched, we finally describe some predicted and/or observed
properties of both helium droplets and trapped condensates
(critical temperature, ground state, excitations, superfluid 
effects, etc.) in order to emphasize relevant analogies and 
differences.

\section{Bose-Einstein condensation}
\label{sec:commonbasis}

The intimate link between helium droplets and trapped condensates
is the occurrence of Bose-Einstein condensation and the fact that
this phenomenon plays a key role in determining most of the properties
of both systems.

In the case of noninteracting particles, Bose-Einstein condensation
has well established features. The system can be represented by a
set of single-particle states of energy $\epsilon_i$, each one occupied
by a certain number of particles $n_i$. At a given temperature $T$,
the average occupation numbers are fixed by  Bose statistics,
resulting in the well known law $\langle n_i \rangle = 1/
\{ \exp [\beta (\epsilon_i - \mu) -1] \} $, where $\beta=
(k_B T)^{-1}$ and $\mu$ is the chemical potential.  The latter is
fixed by the conditions that the total number of particles, $N$, is
equal to the sum $\sum_i \langle n_i \rangle$. For relatively
high $T$ one finds that this condition is satisfied for $\mu <
\epsilon_0$, where $\epsilon_0$ is the energy of the lowest
single-particle state, and the particles are thermally distributed
over many states, all of them having $\langle n_i \rangle$ of the
order of $1$ (or, equivalently, much less than $N$). However, there
is no exclusion principle for bosons: many of them can  stay in
the same state. This happens when the temperature is lowered; when the
chemical potential approaches $\epsilon_0$,  the occupation
of the lowest state becomes a macroscopic number $N_0 \equiv \langle 
n_0 \rangle$ of the order of $N$. This phenomenon is called Bose-Einstein 
condensation.
The critical temperature at which it occurs can be calculated
starting from the knowledge of the single-particle Hamiltonian and
keeping the appropriate thermodynamic limit $N\to \infty$.

In the textbook case of a uniform system of free bosons of mass $m$,
one increases the volume $V$ while keeping the density $n = N/V$
constant; in this way, one gets the critical temperature \cite{huang}
\begin{equation}
k_B T_c =  { 2 \pi \hbar^2 \over m }
\left( {n \over \zeta(3/2)}  \right)^{2/3}
\label{eq:tc1}
\end{equation}
where $\zeta$ is the Riemann zeta function and $\zeta(3/2) \simeq 2.61$.
This result can be rewritten as $n \lambda_T^3 \simeq 2.61$, where
$\lambda_T=[2\pi\hbar^2/ (m k_B T)]^{1/2}$ is the thermal de Broglie
wavelength. This is a quantitative translation of the qualitative
idea mentioned in the introduction: quantum effects become crucial
when the de Broglie wavelength becomes larger than the average
separation between particles. The same calculations also yields the
condensate fraction
$ (N_0/ N ) = 1 - ( T / T_c )^{3/2}$.
At $T=0$ all particles are in the $\epsilon_0=0$ state, that is,
the system is fully condensed. This state has a uniform density and
is characterized by having zero momentum; for this reason, one
often speaks of condensation in momentum space. It is important
to stress here that, when applied to liquid
$^4$He ($m$ and $n$ being the atomic mass and the liquid density,
respectively),  Eq.~(\ref{eq:tc1}) gives a transition temperature
reasonably close to $T_\lambda$, i.e., the temperature at which
helium becomes superfluid. This very simple and striking result
marked the beginning of the theory of superfluids, starting
from the pioneering work of Fritz London \cite{london}.

In the case of $N$ bosons confined in a spherical harmonic potential
$V_{\rm ext}= (1/2) m \omega_{\rm ho}^2 r^2$ , the proper thermodynamic 
limit is
obtained by letting $N\to\infty$ and $\omega_{\rm ho}\to 0$, while
keeping the product $N \omega_{\rm ho}^3$ constant. In this way
the critical temperature is well defined; one finds \cite{bagnato}
\begin{equation}
k_B T_c = \hbar \omega_{\rm ho} \left( {N \over \zeta(3) }
\right)^{1/3} \simeq 0.94 \ \hbar \omega_{\rm ho} \ N^{1/3}
\label{eq:tcho}
\end{equation}
and the condensate fraction is
$( N_0 / N ) = 1 - ( T / T_c )^3$. Notice that the different 
$T$-dependence exhibited by the condensate fraction compared to 
the uniform case is the consequence of the higher density of states
characterizing the harmonic oscillator Hamiltonian.  
At $T=0$ all particles are in the lowest eigenstate of the harmonic
oscillator, namely, $\mu \to \epsilon_0 = (3/2) \hbar
\omega_{\rm ho}$. Thus the particle density has the form of a Gaussian
and the same is true for the momentum distribution. Differently 
from the uniform gas, condensation
occurs in both real and momentum space. Actually, the occurrence
of a sharp peak in the velocity distribution of a cloud of
$^{87}$Rb atoms, released from a magnetic trap, was the
first evidence of Bose-Einstein condensations in these cold
gases \cite{bec95-jila}, and the transition temperature was found to be
very close to the value predicted by Eq.~(\ref{eq:tcho}).

Since both liquid $^4$He and dilute atomic gases are systems of
interacting particles,  a crucial question concerns the 
role played by the interaction, that is, whether and how much the
interatomic forces modify the properties of Bose-Einstein condensation.
Understanding the interplay between quantum statistical and
dynamic effects represents the challenging task to achieve both
experimentally and theoretically in these quantum many-body systems. 

Let us consider a system of $N$ particles interacting {\it via} 
a potential $V({\bf r} - {\bf r}')$ and, possibly, with an 
external field $V_{\rm ext} ({\bf r})$. In principle, one should
solve the exact many-body Schr\"odinger equation or, equivalently,
find the eigenvalues of the Hamiltonian
\begin{equation}
\hat{H}  =  \int \! d{\bf r}  \ \hat\Psi^{\dagger} ({\bf r})
\left[ - {\hbar^2 \over 2m} \nabla^2 + V_{\rm ext}({\bf r}) \right]
\hat\Psi ({\bf r})
 +  {1\over 2} \int \! d{\bf r}'d{\bf r} \
\hat\Psi^{\dagger} ({\bf r}) \hat\Psi^{\dagger} ({\bf r}')
V({\bf r} - {\bf r}')
\hat{\Psi} ({\bf r}') \hat{\Psi} ({\bf r})
\label{eq:manybodyhamiltonian}
\end{equation}
where $\hat{\Psi}({\bf r})$ and $\hat\Psi^{\dagger}({\bf r})$ are
the boson field operators that annihilate and create a particle
at the position ${\bf r}$, respectively. This is a difficult task
and we do not discuss here the possible strategies. We only want
to emphasize the meaning of Bose-Einstein
condensation in terms of a generic solution of the many-body
problem. An elegant way \cite{pinesnoz} consists in writing the
one-body density matrix, which is defined as
\begin{equation}
\rho ({\bf r}',{\bf r}) = \langle \, \hat \Psi ^\dagger
({\bf r}') \hat \Psi ({\bf r}) \, \rangle
\label{eq:densitymatrix}
\end{equation}
where the average is taken in the state that one wants to describe.
This quantity characterizes the correlations between particles
located in different points in space. One can use a complete set
of single particle states to explicitly write $\rho ({\bf r}',{\bf r})$
as a matrix. Then one can diagonalize it, finding eigenstates and
eigenvalues. The latter can be interpreted as the occupation numbers
associated with each eigenstate. In analogy with the ideal Bose gas,
one might expect that below some critical temperature one of these
eigenvalues, instead of being of order $1$ as all the others, is a
number $N_0$ of the order of $N$.  The corresponding eigenstate then
has a special role in the system, since it is macroscopically occupied.
When an interacting system exhibits such a behavior, one says that it
is Bose-Einstein condensed.

In the case of a  uniform system, the appropriate basis is a set
of plane waves, eigenstates of the momentum ${\bf p}$. If
$N_0$ is the number of particles condensed into the lowest
energy ${\bf p}=0$ state, then the density matrix can be written
in the form
\begin{equation}
\rho ({\bf r}',{\bf r}) = N_0 + {\tilde\rho} ({\bf r} - {\bf r}')
\label{eq:rho-uniform}
\end{equation}
with ${\tilde\rho} ({\bf r}-{\bf r}')
= \sum_{\bf p} n_{\bf p} \exp [i {\bf p} \cdot ({\bf r} -
{\bf r}' )/\hbar ]$.  This summation is made over all states
except ${\bf p}=0$; for ${\bf r} - {\bf r}' =0$ it
simply gives the total number of particles with ${\bf p} \ne 0$,
that is ${\tilde\rho} (0) = N -N_0$. In the opposite limit,
$|{\bf r} - {\bf r}'| \to \infty$, due to destructive interference
between the various phase factors, the function $ {\tilde\rho}$
vanishes, while the condensed part of the density matrix remains
everywhere constant. This infinite range of the the density matrix
(off-diagonal long range order) is also called first order coherence.
It is a key consequence of Bose statistics 
and is the origin of the quantum correlations that are responsible for
superfluidity. This is the basic physics behind the superfluid behavior
of liquid $^4$He below $T_\lambda$. In such a dense system,  the
effect of the interaction, which is implicitly included in
Eq.~(\ref{eq:rho-uniform}), is very important.  Compared to
the ideal gas, it changes the features of the phase transition
and gives a large depletion of the condensate, $N_0/N$ being of
the order of $10$\%  even at $T=0$.

This description can be easily generalized to nonuniform systems.
In this case, the contribution to $\rho ({\bf r}',{\bf r})$ associated 
with the condensed state
is no more a number, but depends on positions. The straightforward
generalization of Eq.~(\ref{eq:rho-uniform}) is
\begin{equation}
\rho ({\bf r}',{\bf r}) =  \Phi^*({\bf r}) \Phi({\bf r}') +
{\tilde\rho} ({\bf r},{\bf r}')
\label{eq:rho-nonuniform}
\end{equation}
where $\Phi^*({\bf r}) \Phi({\bf r}')$ is of order $N$, while
${\tilde\rho} ({\bf r},{\bf r}')$ vanishes for large
$|{\bf r} - {\bf r}'|$ \cite{penrose}. This is
equivalent to say that the bosonic field operator splits in two
parts
\begin{equation}
\hat \Psi ({\bf r}) = \Phi ({\bf r}) + \hat\Psi' ({\bf r}) \, ,
\label{eq:decomposition}
\end{equation}
where the first term is a complex function associated with the
macroscopic occupation of the condensate and the second one
is the field operator associated with the noncondensed particles.
The complex function can be defined as the expectation value of the 
field operator:
\begin{equation}
\langle \hat{\Psi} \rangle  \equiv  \Phi = | \Phi | e^{iS} \; . 
\label{eq:orderparameter}
\end{equation}
It behaves as a classical field having the meaning of an order parameter, 
and is often named {\sl macroscopic wave function}. Its modulus 
gives the condensate density through $n_0({\bf r})=|\Phi({\bf r})|^2$,
while the phase $S$ can be used to define a velocity field through
${\bf v} = (\hbar / m)  \mbox{\boldmath$\nabla$} S$. 

Similarly to the case of uniform gases, the fact that the order 
parameter has a well-defined phase corresponds to assuming the 
occurrence of a broken gauge symmetry in the
many-body system. It is worth noticing that, strictly speaking, in a
finite-sized system neither the concept of broken gauge symmetry,
nor the one of off-diagonal long-range order can be applied. In fact,
the density matrix (\ref{eq:rho-nonuniform}) vanishes for distances 
of the order of the size of the system so that, in order to make the
separation of the condensate and noncondensate components physically
meaningful, it is crucial that the lengthscale at which ${\tilde\rho} 
({\bf r},{\bf r}')$ vanishes be much smaller than the size of the 
system.  The
condensate wave function $\Phi$ can nevertheless be always calculated 
as the eigenfunction of the density matrix with the largest eigenvalue.  
A numerical  diagonalization of the density matrix was performed, for 
example,  by  Lewart {\it et al.} \cite{lewart} in order to estimate the 
condensate density, $n_0({\bf r})$, in helium droplets with up to $240$ 
atoms, using a variational Monte Carlo approach.

\section{Dilute gases and Gross-Pitaevskii theory}
\label{sec:dilute}

The decomposition of the field operator in Eq.~(\ref{eq:decomposition})
is particularly useful when the depletion of the condensate, i.e., the
fraction of noncondensed particles, is very small. This happens when
the interaction is weak, but also for particles with arbitrary
interaction, provided the gas is dilute. In this case, one can expand
the Hamiltonian (\ref{eq:manybodyhamiltonian}) by treating $\hat\Psi'$
as a small quantity.

In a uniform gas the condensate wave function, $\Phi$, is just a
constant and expanding $\hat H$ to the lowest order in $\hat\Psi'$ gives 
access to the equations
for the excited states. This procedure was introduced by Bogoliubov
\cite{bogoliubov}, who found an elegant way to diagonalize the
Hamiltonian by using simple linear combinations of particle
creation and annihilation operators. These are known as Bogoliubov's
transformations and stay at the basis of the concept of
{\sl quasiparticle}, which is one of the most important concepts
in quantum many-body theory.

If the gas is nonuniform, the theory has to include an equation for
the spatial variation of the condensate, even at the zeroth-order in
$\hat\Psi'$. A possible strategy consists in writing the Heisenberg
equation for the evolution of the field operators, using the Hamiltonian
(\ref{eq:manybodyhamiltonian}):
\begin{equation}
i\hbar \frac{\partial}{\partial t} \hat \Psi({\bf r},t) = 
[ \hat \Psi, \hat H ] 
=  \left( -\frac{\hbar^2\nabla ^2}{2m} +
V_{\rm ext}({\bf r})  +\int d{\bf r}' \
\hat \Psi ^{\dagger}({\bf r}',t) V({\bf r}'-{\bf r})
\hat \Psi ({\bf r}',t) \right) \hat \Psi ({\bf r},t) \; .
\label{eq:H}
\end{equation}
Then the lowest order is obtained by replacing the operator $\hat\Psi$
with the classical field $\Phi$. In the integral containing the atom-atom
interaction $V({\bf r}'-{\bf r})$, this replacement is, in general, a poor
approximation when short distances $({\bf r}'-{\bf r})$ are involved. In a
dilute and cold gas, one can nevertheless obtain a proper expression for the
interaction term by observing that, in this case, only binary collisions at
low energy are relevant and these collisions are characterized by a single
parameter, the $s$-wave scattering length, independently of the details of
the two-body potential. This allows one to replace $V({\bf r}'-{\bf r})$ in
(\ref{eq:H}) with an effective interaction $V({\bf r}'-{\bf r}) = g \delta
({\bf r}'-{\bf r})$ where the coupling constant $g$ is related to the
scattering length $a$ through  $g =  4 \pi \hbar^2 a /m$.
Using this pseudo-potential in Eq.~(\ref{eq:H}) and replacing
$\hat\Psi$ with $\Phi$, one gets the following closed equation
for the order parameter:
\begin{equation}
i \hbar { \partial \over \partial t} \Phi ({\bf r},t) =
 \left( - { \hbar^2 \nabla^2 \over 2m } + V_{\rm ext}({\bf r})
+g |\Phi({\bf r},t)|^2 \right) \Phi ({\bf r},t) \; .
\label{eq:TDGP}
\end{equation}
This is known as Gross-Pitaevskii (GP) equation \cite{gross,pitaevskii}.
It has been derived assuming that $N$ is large while the fraction of
noncondensed atoms is small. On the one hand, this means that quantum
fluctuations of the field operator have to be small, which is true when 
$n|a|^3 \ll 1$, where $n$ is the particle density. In fact, one can
show that, at $T=0$ the depletion of the condensate is proportional
to $(n |a|^3)^{1/2}$ \cite{huang}. On the other hand, thermal
fluctuations have also to be negligible and this means that the theory
is limited to temperatures much lower than $T_c$. Within these limits,
one can identify the total density $n$ with the condensate 
density $n_0$. 

The stationary solution of Eq.~(\ref{eq:TDGP}) corresponds to the
condensate wave function in the ground state. One can write
$\Phi({\bf r},t)=\phi({\bf r}) \exp(-i \mu t/ \hbar)$,  where
$\mu$ is the chemical potential and $\phi$ is real and normalized
to the total number of particles, $\int d{\bf r} \ \phi^2
= N_0 = N$.  Then the GP equation becomes
\begin{equation}
\left( - { \hbar^2\nabla^2 \over 2m } + V_{\rm ext}({\bf r})  +
g \phi^2 ({\bf r}) \right) \phi({\bf r}) = \mu  \phi({\bf r}) \; .
\label{eq:GP}
\end{equation}
This has  the form of a ``nonlinear Schr\"odinger equation", the
nonlinearity coming from the mean-field term, proportional to the
particle density $n({\bf r})=\phi^2({\bf r})$. It is worth noticing
that the same equation can be obtained by minimizing the energy
of the system written as a functional of the density:  
\begin{equation}
E[n] \  = \  \int d{\bf r} \  \left[ {\hbar^2 \over 2m}
| \mbox{\boldmath$\nabla$} \sqrt{n} |^2
+  n V_{\rm ext}({\bf r})  + { g n^2 \over 2} \right]
\ = \ E_{\rm qp} + E_{\rm ext} + E_{\rm int}
 \; .
\label{eq:funcofdensity}
\end{equation}  
The first term corresponds to the quantum kinetic energy coming 
from the uncertainty principle; it is usually named ``quantum 
pressure" and  vanishes for uniform systems. 

In the absence of interactions ($g=0$), the GP equation (\ref{eq:GP})
reduces to the usual single-particle Schr\"odinger equation. Conversely,
when $g \ne 0$, the mean-field contribution can have an important role 
in determining the energy and density distribution of the condensate, 
despite the smallness of the gas parameter $n|a|^3$. In fact, what 
matters is the relative weight of the mean-field potential, the kinetic 
energy, and the external potential. As we will see later on, one
can easily find situations where the gas is dilute but strongly 
{\it nonideal}.  It is also worth pointing out the key role 
played by the chemical potential  in the GP theory, where the 
time dependence of the stationary order parameter is fixed by $\mu$ 
and not by the energy. This is a consequence of the fact that
the order parameter is not a wave function and that the GP equation 
is not a Schr\"odinger equation in the usual sense of quantum mechanics. 
From the point of view of many-body theory the order parameter 
corresponds to the matrix element of the field operator between 
two many-body wave functions containing, respectively,  $N$ and $N+1$ 
particles. This implies that its  time dependence is fixed by the 
factor $\exp \{ -i[E(N+1) -E(N)]t \}$ and hence by the chemical potential 
$\mu=\partial E/\partial N$ rather than by the energy $E$. 

The excited states of the condensate, at $T=0$, can be also
calculated starting form the time dependent GP equation
(\ref{eq:TDGP}), by looking for small deviations around the
ground state in the form
\begin{equation}
\Phi ({\bf r},t) = e^{-i\mu t/\hbar} \left[ \phi ({\bf r}) +
u({\bf r}) e^{-i \omega t} + v^*({\bf r}) e^{i \omega t} \right]
\; .
\label{eq:linearized}
\end{equation}
By keeping  terms linear in the complex
functions $u$ and $v$, Eq.~(\ref{eq:TDGP}) becomes
\begin{eqnarray}
 \hbar \omega u({\bf r}) &=& [ H_0 - \mu + 2 g \phi^2({\bf r})]
u ({\bf r}) + g  \phi^2({\bf r}) v ({\bf r})
\label{eq:coupled1}
\\
- \hbar \omega v({\bf r}) &=& [ H_0 - \mu + 2 g \phi^2({\bf r})]
v ({\bf r}) + g  \phi^2({\bf r}) u ({\bf r}) \; .
\label{eq:coupled2}
\end{eqnarray}
where $H_0= - (\hbar^2/2m) \nabla^2 +  V_{\rm ext}({\bf r})$. These
coupled equations allow one to calculate the eigenfrequencies $\omega$
and hence the energies $\varepsilon= \hbar \omega$ of the excitations.
In a uniform gas, the amplitudes $u$ and $v$ are plane waves and
one recovers the Bogoliubov's spectrum \cite{bogoliubov}
\begin{equation}
(\hbar \omega)^2 = \left( {\hbar ^2 q^2 \over 2m} \right) \left(
{ \hbar^2 q^2 \over 2 m} + 2 g n \right)
\label{eq:bogoliubovspectrum}
\end{equation}
where ${\bf q}$ is the wavevector of the excitations. For large
momenta the spectrum coincides with the free-particle energy
$\hbar^2q^2/2m$. At low momenta Eq.~(\ref{eq:bogoliubovspectrum})
instead gives the phonon dispersion $\omega=cq$, where $ c = [g 
n /m]^{1/2}$ is the sound velocity. This transition between
the two regimes occurs when the typical wavelength of the excited 
states becomes  of the order of the so called {\sl healing length},
\begin{equation}
\xi = [8\pi n a ]^{-1/2} \; , 
\label{eq:xi}
\end{equation}
which is a rather important lengthscale for superfluidity. When
the order parameter is forced to vanish at some point (by an 
impurity, a wall, or something else), the healing length is the
typical distance over which it recovers its bulk value. In a 
nonuniform condensate the lowest excitations are not plane waves, 
but they have still a phonon-like character, in the sense that they
involve a collective motion of the condensate, and the transition
from phonon-like to single-particle excitations is still 
an important feature of these systems. 

An important point which deserves to be stressed is the existence 
of a close connection between the dynamics of a condensate, expressed 
by the GP equation, and the hydrodynamic equations for an irrotational 
and viscousless fluid. The natural variables for the hydrodynamic 
description are density and velocity field, which can be related
to the modulus and the phase of the order parameter as in 
Eq.~(\ref{eq:orderparameter}). Using these definitions into the GP
equation (\ref{eq:TDGP}) and neglecting the contribution of the
quantum pressure, $\nabla ^{2} \sqrt{n}$, one gets the two coupled 
equations  
\begin{eqnarray}
 { \partial \over \partial t}  n &+&  \mbox{\boldmath$\nabla$}
\cdot ({\bf v}n) = 0
\label{eq:continuity} \\
m  { \partial \over \partial t}  {\bf v} &+&
\mbox{\boldmath$\nabla$} \left(  \mu  + { mv^2 \over 2 } \right) 
= 0 
\label{eq:euler}
\end{eqnarray}
where $\mu ({\bf r},t) = g n ({\bf r},t) + V_{\rm ext}({\bf r})$.  
These are the continuity and Euler equations for an irrotational
and viscousless fluid in the collisionless regime, that is, the 
hydrodynamic equations for a superfluid at $T=0$. Neglecting the 
quantum pressure term in the GP equations means that 
Eqs.~(\ref{eq:continuity}) and (\ref{eq:euler}) are valid for 
{\it macroscopic} motions of the condensate, i.e., excitations
with wavelength larger than the healing length. 
  
The equations written in this section stay at the basis 
of most of the theoretical
papers recently written on BEC in trapped gases (see \cite{rmp}
for a recent review). Analytical and numerical methods have been
developed in order to solve both the stationary and time-dependent
Gross-Pitaevskii equation and the results are often in excellent
agreement with experiments. On the other
hand, the accuracy of the approximations made in deriving the
GP equation can be directly tested by performing Monte Carlo
simulations, which are exact within statistical errors. For trapped
condensates, this has been done in \cite{krauth}
using a Path Integral Monte Carlo method, showing that the
stationary GP
equation is very accurate for $T$ less than about $0.75 T_c$
and providing quantitative estimate for the effects of correlations
beyond mean-field theory. Such corrections has been also
calculated for a uniform hard-sphere Bose gas in \cite{giorgini}
using a Green Functions Monte Carlo method.

It is worth mentioning that the theory of weakly interacting Bose
gases born well before the observation of BEC in trapped gases,
as an attempt to explain the behavior of superfluid helium. In the
case of helium, however, the parameter $n|a|^3$ is of the order
of $1$, so that the theory looses most of its
predictive power.

\section{Dense fluids and the quantum many-body problem}
\label{sec:dense}

When the system is neither weakly interacting nor dilute, the
mean-field theory based on Eq.~(\ref{eq:TDGP}) is no more
applicable. Quantum correlations beyond mean-field are
essential; they can give, for instance, a large depletion of the
condensate even at $T=0$, so that perturbative schemes based on
the smallness of $\hat\Psi'$ in Eq.~(\ref{eq:decomposition}) fail. 
In order to overcome this obstacle, one can follow
two alternative and complementary paths: i) keep the theory
at a {\sl microscopic} level by solving the many-body problem
as accurately as possible; ii) develop phenomenological theories,
starting from a more {\sl macroscopic} view of the liquid. 

Among the microscopic approaches one can distinguish different
strategies depending on the use or non-use of stochastic Monte Carlo 
procedures. Example of non-stochastic approaches are variational 
methods and perturbative schemes. Most of them relies on diagram 
expansion and summation techniques typical of quantum many-body
theories. Jastrow-Feenberg \cite{feenberg} wave functions are often 
used in this context as a starting point. They can be combined with 
integral equations as in the optimized hypernetted chain (HNC)
theory or with perturbative schemes as in the correlated basis 
function (CBF) theory. The energy minimization within a certain 
class of variational wave functions can be performed with stochastic
techniques as in the variational Monte Carlo (VMC) method. This
allows one to include more sophisticated correlations in the trial 
wave function  as, for instance, in the recent Shadow Wave Function 
theory. Monte Carlo algorithms can finally be used to get exact 
results with statistical uncertainties.  This is the case of approaches 
like Green's function MC, Diffusion MC and Path Integral MC, which are 
powerful, reliable, and computationally demanding. It is not
our aim to present an overview of all these theories and calculations.
Appropriate references can be found in the various papers where 
these techniques are applied to helium droplets (see, for instance, 
\cite{microscopic} and references therein).

Conversely, a problem that deserves to be mentioned at this point is
whether it is possible to make a bridge between dilute gases and dense 
fluids, like helium, within a single theoretical scheme. The recent 
Variational Monte Carlo  calculation by DuBois and Glyde \cite{glyde}
is an interesting effort in this direction. By using a rather simple 
variational wave function,  they evaluated the ground state properties 
of a trapped Bose gas made of hard-sphere of radius $a$, over a
wide range of the parameter $na^3$. In particular, through the
diagonalization of the one-body density matrix, they calculated the
condensate fraction showing how it decreases from $1$ to about $10$\%
in bulk, when $na^3$ goes from $0$ to the value appropriate for
liquid helium, which is about $0.2$-$0.3$ (the range of the
interaction is about $2$-$3$ \AA, while the density is
approximately $0.022$\AA$^{-3}$). They also pointed out interesting
features about its spatial dependence, i.e., the tendency for dense
fluids to have larger condensate fraction at the surface where the
density is smaller. This enhancement of the condensate was first
found in the Variational MC calculations of Ref.~\cite{lewart} in
the case of helium droplets.

Phenomenological approaches can also be applied to study dense 
systems. For confined fluids the simplest one is the liquid drop 
model. One considers a droplet of constant density within a radius 
$R$ and a sharp surface. Then one can write the energy as a power 
expansion in $R$ (or, equivalently, 
in power on the number of particles $N$) taking bulk energy, surface 
energy, curvature energy, etc., as input parameters. In the limit of 
large $N$, one can consider the dynamics of these droplets as due
to large wavelength, collective motions. Then, it makes sense to use 
hydrodynamic equations with appropriate boundary conditions in 
order to get the dispersion of bulk (phonon-like) and surface 
(ripplon-like) excitations \cite{varenna,casas}. Several properties 
of quantum fluid droplets are well predicted by this type of 
models \cite{bohr};  what is missing is the accurate description of 
structures and excitations on a more microscopic scale, i.e., at 
the level of atomic distances. A step further in
this direction is the density functional approach. 

In density functional theories at zero temperature, the energy of 
a system is assumed to be a functional of the particle density 
$n ({\bf r})$. A good starting point for Bose fluid is given by  
\begin{equation}
E[n] \ =  \ E_c[n] +  \int d{\bf r} \  
\left[ {\hbar^2 \over 2m} | \mbox{\boldmath$\nabla$} \sqrt{n} |^2  
+  n V_{\rm ext}({\bf r}) \right]  \; , 
\label{eq:ec}
\end{equation}
which is the natural generalization of the Gross-Pitaevskii functional 
(\ref{eq:funcofdensity}).  Ground state configurations are obtained 
by minimizing this energy with respect to the density. This leads to 
the Hartree-type equation 
\begin{equation}                                                     
\left( - { \hbar^2\nabla^2 \over 2m } + V_{\rm ext}({\bf r})  
+ U [n({\bf r})] \right)
\sqrt{n({\bf r})} \ = \ \mu \sqrt{n({\bf r})} \ \ \ ,
\label{eq:hartree}
\end{equation}
where $U[n] \equiv \delta E_c /\delta n({\bf r})$ acts as a  
mean field,  while the chemical potential $\mu$ is introduced in order 
to ensure the  proper normalization  of the density to a fixed number 
of particles. Equation~(\ref{eq:hartree}) exhibits a formal analogy
with the Gross-Pitaevskii equation (\ref{eq:GP}). The main conceptual 
difference is that Eq.~(\ref{eq:hartree}) is an equation for the
density and not for the order parameter. While in a dilute gas setting 
$n=|\Phi|^2$ is an excellent approximation, in a correlated system it
is meaningless. 

The correlation energy $E_c[n]$ incorporates the effects of dynamic 
correlations induced by the interaction. The fact that this energy is
a unique functional of $n({\bf r})$ is ensured by the Hohenberg-Kohn 
theorem \cite{kohn}.  In the case of dilute gases, one simply has 
$E_c[n]= \int d{\bf r} (1/2) gn^2$, as in Eq.~(\ref{eq:funcofdensity}).  
Conversely, since liquid helium is a strongly correlated system, a 
rigorous derivation of $E_c[n]$, starting from first principles, is 
not available. One  then resorts to approximate  schemes for the 
correlation energy (see, for instance, Refs.~\cite{kro,griffin} for 
discussions of this problem from the viewpoint of microscopic
many-body theory). A simple but useful approach consists of writing  
a phenomenological expression for the  correlation energy, whose  
parameters are fixed to reproduce  known  properties of the bulk liquid.
A functional of this type was introduced in Ref.~\cite{Str87a,Str87b} to
investigate properties of the free surface and droplets of both $^4$He  
and $^3$He. The correlation energy was written as
\begin{equation}
E_c[n] = \int d{\bf r}  \left[ {b\over 2} n^2 + {c\over 2} 
n^{2+\gamma} + d (\nabla n)^2 \right] \ \ \ ,
\label{eq:skyrme}
\end{equation}
where $b,c$ and $\gamma$ are phenomenological parameters fixed   to 
reproduce the ground state energy, density and compressibility of the
homogeneous liquid at zero pressure, and $d$ is adjusted to the surface 
tension of the liquid. The first two terms correspond  to a local
density  approximation  for the correlation energy, while nonlocal
effects  are included through the gradient correction. Non-locality 
effects have been included in DFT in a more realistic  way by 
Dupont-Roc et al. \cite{Dup90}, who generalized  Eq.~(\ref{eq:skyrme}) 
to account for the finite range of the atom-atom interaction. Further 
improvements has been added in  Ref.~\cite{ot}. These and other similar
functionals have been used for calculations of ground state properties 
of liquid helium in different geometries (see \cite{dft-papers} and
references therein). 
 
Dynamical problems can also be faced with density functionals.
On the one hand, one can include in the functional terms which
explicitly depends on the current density, as done in \cite{ot},
thus obtaining two coupled equations for both density and velocity 
field, which can be numerically solved. Excited states of helium 
droplets have been obtained in this way \cite{palma,hernandez};
the linearized equations of motions, in this case,  correspond
to a generalization of the Bogoliubov equations
(\ref{eq:coupled1})-(\ref{eq:coupled2}), formally equivalent to the
equations of the Random Phase Approximation (RPA) \cite{pinesnoz}
with an effective interaction of phenomenological nature. On the other
hand, one can simplify the problem by using a purely hydrodynamic
approach, in which the equations of motions for the velocity field
are the ones of an irrotational and viscousless fluid, but with the
density taken from the minimization of the density functional
(\ref{eq:ec}). This viewpoint was chosen, for instance, in
Ref.~\cite{princeton} to predict the moment of inertia
of doped helium droplets.

\section{Like and unlike features of helium droplets and
trapped condensates}
\label{sec:features}

\subsection{Temperature}

In typical experiments the temperature of $^4$He droplets is about 
$0.37$ K \cite{hartmann-T},  as the result
of a spontaneous evaporation of the warmest atoms after the formation of
droplets in the free jet expansion of helium \cite{brink}. This
temperature is well below the bulk superfluid transition temperature,
$T_\lambda= 2.17$ K. Path Integral Monte Carlo calculations \cite{sind}
have shown that even small droplets, with few tens of atoms, are superfluid
at this temperature. Moreover, the droplet is cold enough to neglect the
thermal activation of excited states. The lowest ones are discretized
collective modes (bulk and surface oscillations) which are not 
significantly populated at $T \sim 0.4$ K. Thus, for most purposes, helium 
droplets can be considered as $T=0$ systems.

Trapped condensates can be cooled down to few tens of nK. Again this
is the result of evaporation, but here the evaporative cooling process 
is induced and controlled through an external rf-field which acts on the 
shape of the confining potential, lowering the edge of the trap and 
letting the hottest atoms to escape. This cooling process can be 
continued until most of the thermally excited atoms are removed and, 
hence, the system can be considered as a condensate at $T=0$. 
Differently from helium droplets,  the properties of trapped gases can 
also be studied as a function of $T$, just stopping the evaporative 
cooling at the desired temperature. In this case, the gas has two 
components, a condensate and a thermal cloud, and the Gross-Pitaevskii 
theory of section \ref{sec:dilute} has to be generalized to properly 
include the thermal component. In the following, however,  we restrict 
the analysis to $T=0$ \cite{noteT}.

\subsection{Ground state}

Let us consider a typical helium cluster of $10^3$-$10^4$ atoms. In
its ground state, it is a self-bound liquid droplet with a rather flat 
density distribution, the central density being close to the one of 
the uniform liquid in the limit of zero pressure, i.e.,  $n= 0.022$ 
\AA$^{-3}$ (see Fig.1). Assuming the droplet to have constant density 
and sharp surface, its radius would be $R = r_0 N^{1/3}$ with $r_0 \simeq 
2.2$ \AA, which gives about $20$ to $50$ \AA, significantly 
larger than the average atomic distance. Actually, the true density 
profile has a rather smooth surface due to the large zero point motion 
of the atoms. The density decreases from the bulk value to zero within
about twice the average atomic distance, i.e., about 6 to 9 \AA\
depending on the precise definition of the surface thickness (see
\cite{harms} and references therein). The thickness of the surface 
region is comparable to the droplet radius for $N$ of the order
of $100$ or less, but even for larger droplets the diffuseness of the
surface can play a significant role. For instance, it makes the
average density of the droplet lower than the bulk liquid value and
hence the  atoms are less bound to the droplet than to the uniform
liquid. As already said, one can reasonably use the liquid drop model 
to express the energy per particle as a function of $N$:
\begin{equation} 
{E \over N} = a_v + a_s N^{-1/3} + a_c N^{-2/3} + \dots
\label{eq:energydrop} 
\end{equation}
where $a_v \simeq -7.2$ K is the volume coefficient, i.e. the energy per
particle in the bulk liquid (in the limit of zero temperature and
zero pressure), and the next terms are surface and curvature energies.
This type of formulae are frequently used to fit the numerical 
results of ground state calculations.

The density distribution of a pure helium droplet is determined by the
balance of the kinetic energy, associated with the zero-point motion
of the atoms, and the potential energy due to atom-atom interaction.
One can also produce density modulations in the droplets by picking up 
a foreign atom or molecule, acting as an external potential in 
which helium atoms readapt. This is, however, a local effect, in the 
sense that the main changes in the density distribution are limited 
to the first layers of atoms around the impurity. 

Differently from helium droplets, trapped condensates are not
self-bound.  They are produced and maintained in an external confining
harmonic potential, of the form $V_{\rm ext}= (1/2) m \omega_{\rm ho}^2
r^2$ \cite{note_anis}. If this potential is switched-off, the condensate
freely expands. 

The density distribution of a trapped gas is determined by the balance
of three energy contributions given in Eq.~(\ref{eq:funcofdensity}).
Since the sign of the interaction energy is fixed by the scattering
length $a$, one may have either a repulsive or attractive mean-field
potential, for positive and negative $a$, respectively. The case of
negative $a$ is very interesting for several reasons but it is not
suitable for a comparison with helium. In fact, a uniform gas of
atoms with negative $a$ is unstable; it has negative pressure and
it lowers its energy by collapsing into separate regions of high
density. In a trap, quantum pressure can prevent this collapse, at
least below some critical density \cite{rmp}, but such metastable
condensates have no counterpart in the uniform limit, differently
from the case of liquid helium and of trapped condensates with
positive $a$.

If the atoms were not interacting ($a=0$) the condensate would have
the form of the lowest single-particle state in the harmonic
potential, i.e, a Gaussian having a width $a_{\rm ho} = [ \hbar
/ (m\omega_{\rm ho}) ]^{1/2}$. In actual condensates, with $a>0$,
the repulsion between atoms tends to lower the central density,
thus increasing the condensate width. The equilibrium
configuration is again fixed by the balance between $E_{\rm qp}$,
$E_{\rm ext}$ and $E_{\rm int}$. It turns out that for $Na/a_{\rm ho}
\gg 1$ (Thomas-Fermi limit) the density in the central part is so
smooth that the quantum pressure is negligible compared to both
$E_{\rm ext}$ and $E_{\rm int}$. In this case, the GP equation
(\ref{eq:GP}) simply gives $n(r)=g^{-1} [ \mu - V_{\rm ext} (r)]$,
that is, a density having the form of an inverted parabola. 
In the same limit, one can easily calculate the energy per particle, 
\begin{equation} 
{E \over N} =  {5 \hbar \omega_{\rm ho} \over 14}
\left( { 15 N a \over a_{\rm ho} } \right)^{2/5} \; .
\label{eq:energytf} 
\end{equation}
which contains both the harmonic potential parameters and the 
scattering length. Only in a narrow region at the surface, where 
the density vanishes, the interaction term becomes small and the 
density profile is determined by the balance between quantum pressure 
and external potential \cite{dps}; it deviates from the inverted parabola 
and goes smoothly to zero. This surface region gives a logarithmic 
correction to the Thomas-Fermi energy (\ref{eq:energytf}), of 
the form $E/N \sim N^{-2/5} \ln N$. An example is shown in Fig.1,
where one can see how the external potential affect the overall 
shape of the condensate (right panel), in contrast with the 
self-bound helium droplet (left panel).    

At present, several types of condensates are available in different
labs (different geometries and atomic species). Let us consider a
common situation of a condensate made by $10^5$-$10^6$ atoms
of sodium or rubidium, in a trap with $\omega_{\rm ho}/(2\pi) \sim
10$-$100$ Hz. The width of the condensate can 
be several $\mu$m, but wider condensates, up to $300 \mu$m, have 
been obtained in anisotropic traps. The scattering length $a$ is of 
the order of a few nm and the ratio $a/a_{\rm ho}$ is about $10^{-3}$, 
so that $Na/a_{\rm ho} \gg 1$; thus the density profile is very well 
represented by an inverted parabola with a thin surface region. The size 
of the system is much larger than the average interatomic distance, 
of the order of $10^{-7}$ m, as happens for large helium droplets.  
The central density is of the order of $10^{13}$-$10^{15}$ cm$^{-3}$, 
so that $na^3$ is less than $10^{-3}$, which ensures the applicability 
of Gross-Pitaevskii theory. 

The relevant lengthscales are summarized in Table I for a $^4$He droplet 
with $10^4$ atoms and a trapped condensate of $10^5$ atoms of $^{87}$Rb. 
One notes that the only difference in $d$, $\xi$ and $R$  is a common 
rescaling, from nm (helium) to $\mu$m (trapped BEC). This is not true 
for $a$, which is of the same order of $d$ in the dense fluid and much 
smaller than $d$ in the dilute gas. 

A direct consequence of the very different values of $na^3$ for the
two systems in Fig.1 is that the total density plotted in the right 
panel can be also identified with the condensate density, since quantum 
depletion is negligible (less than $1$\%), while the same is 
not true for helium. In the latter case, the condensate density is 
expected to be about $10$\% of the total density in the inner part
of the droplet, increasing to a maximum value at the surface where
the system is more dilute \cite{lewart} (see also \cite{galli} for 
a recent discussion).

\subsection{Excitations}

A peculiar feature of Bose superfluids is that their excitations at 
low energy correspond to collective modes, which can be described as 
fluctuations of the order parameter. This happens when their wavelength 
is larger than the healing length. 

For uniform and dilute gases, the spectrum of excitations is given 
by the Bogoliubov result (\ref{eq:bogoliubovspectrum}), which is
phononic at low $q$ and single-particle at high $q$. Low $q$ 
excitations are phonons even in liquid helium, with sound velocity 
$c \sim 238$ m/s. However, since the fluid is dense and highly correlated, 
the interpolation between the phononic and single-particle regimes 
is more subtle. The phonon branch reaches a maximum at $q \sim 
1$\AA$^{-1}$ and then forms a rather deep minimum at $q \sim 
1.9$\AA$^{-1}$, with energy  $\sim 8.7$ K. The excitations near this 
minimum, whose dispersion  is approximately parabolic, are called 
rotons. Their wavelength is of the order of the interatomic distances 
and slightly larger than the healing length, so that they still have  
collective character, but are related, roughly speaking, to 
a tendency to local order on the atomic scale. 
 
Helium droplets and trapped BEC have a finite radius $R$. As a 
consequence,  the spectrum is discrete and the excited states have
to be classified according to the number of radial nodes, $n_r$, and the 
angular momentum quanta ($l$ and $m$, if the system is spherical). 
The discretization is particularly important for the lowest energy 
excitations, whose wavelength is comparable with $R$, and   
correspond to oscillations of the whole system.  

An example of collective mode is the state with $n_r=1$ and $l=m=0$, 
which is a purely compressional oscillation (monopole, or {\it breathing} 
mode). By keeping $l=m=0$ and increasing 
$n_r$, one finds density oscillations in the radial direction having 
wavelength smaller than $R$. If $n_r$ is large, so that the wavelength 
becomes much smaller than $R$ but still larger than $\xi$ (see Table I),
these modes can be thought as stationary states of bulk excitations 
(phonons and/or rotons) which propagate radially and reflect at the 
surface. For large $R$, these discretized states approach a continuum
and the spectrum becomes closer and closer to the one of a uniform 
system, namely, the phononic Bogoliubov dispersion in dilute gas and 
the phonon-roton branch in liquid helium. 

Another interesting class of excitations is the one with $n_r=0$ and 
$l \ge 2 $. These modes correspond to shape oscillations, or surface 
excitations, because the motion is mainly concentrated near the surface, 
the latter being displaced in and out periodically. The lowest one is  
the quadrupole deformation, $l=2$. By increasing $l$ one gets shape 
oscillations with more nodes in the angular direction, that is, 
surface waves with shorter wavelength. In the absence of external 
confinement as in helium droplets, and for large $R$, these states 
approach the collective waves of the liquid-vapour interface (ripplons), 
having dispersion $\omega^2 = q^3 \sigma / (m n)$, where 
$\sigma$ is the surface energy and $q$ is the wavevector parallel 
to the surface. In the case of trapped gases the dispersion is instead 
given by the law $\omega^2 = q F/m$, where $F=m\omega_{\rm ho} R$ is the 
trapping harmonic force calculated at $r=R$ \cite{alkawaja}.

How can these collective states be excited and observed? In trapped
condensate the answer is simple: one can modulate the external potential
in order to force oscillations in the system in a classical way. With this 
method,  one has 
already observed monopole, dipole, quadrupole modes, as well as higher 
multipolarities \cite{varenna1,varenna2,MIT-surface}. Their frequency 
is measured with great accuracy  and the agreement with theory is very 
good. The theoretical work  basically corresponds to the numerical solution 
of the Bogoliubov-like equations (\ref{eq:coupled1})-(\ref{eq:coupled2}). 
In  the Thomas-Fermi limit,  $Na/a_{\rm ho} \gg 1$, one can also 
obtain analytic results by linearizing the hydrodynamic equations 
(\ref{eq:continuity})-(\ref{eq:euler}) \cite{stringari-hd}. In 
spherical traps one obtains the dispersion relation $\omega (n_r,
\ell) = \omega_{\rm ho} (2n_r^2 + 2n_r\ell + 3n_r +\ell)^{1/2}$, which 
can be compared with the spectrum of noninteracting particles in the 
same harmonic potential, $\omega (n_r,\ell) = \omega_{\rm ho} (2n_r + 
l)$ \cite{note-dipole}. 

Phonon-like excitations with wavelength smaller than the condensate 
size can also be produced. For instance,
one can suddenly switch-on a narrow laser beam, focused in the center 
of the trap. An optical dipole force acts on the atoms generating a 
wavepacket of excitations which then moves throughout the condensate as 
a sound wave. The velocity of this sound wave has been measured in  
\cite{MIT-sound}, finding good agreement with the prediction of 
Bogoliubov theory. Phonon-like excitations have been also generated
in light-scattering experiments \cite{MIT-bragg}.

Driving collective excitations in helium droplets, in a controllable 
way, is much less simple, since there is no confining potential 
to play with. However, one can produce and probe excitations
by picking up impurities. A foreign atom or molecule, laying inside
the droplet or at the surface, can be excited with a laser and this 
excitation may couple to the internal degrees of freedom of the 
droplet. This mechanism is at the basis of the so-called depletion 
method: a photon absorbed by an embedded impurity causes a transfer 
of energy from the impurity to the droplet {\it via} the creation 
of excitations. The latter eventually exchange their energy with 
atoms at the surface, letting them to evaporate through a quantum 
evaporation process. Thus, by monitoring the droplet beam with a mass 
spectrometer, one detects the spectroscopic transitions of the 
impurity which, in turn, contain information about the droplet 
itself \cite{scoles92}. This process has some interesting 
analogies with the quantum evaporation experiments in bulk helium 
\cite{quantumevap}, where the initial bunch of excitations (phonons
and/or rotons) is produced by a heater immersed in the liquid, 
while the evaporated atoms are collected at a bolometer above the
surface. There are also significant analogies with the sound
propagation experiment in a trapped condensate \cite{MIT-sound}, where 
the  ``heater'' is represented by the laser beam. In this case, if the 
trap is kept on, phonons do not evaporate atoms, but they may exchange 
energy with single-particle states in the outer part of the condensate. 

An example of how the superfluid environment affects the spectrum
of embedded impurities is the observation of the phonon wing in the
electronic excitation spectrum of a glyoxal  (C$_2$H$_2$O$_2$) 
molecule in helium droplets with about $5500$ atoms \cite{hartmann}.
The observed features were found to be consistent with the 
existence of a phonon-roton spectrum of excitations inside the liquid
droplet, with no low-lying single-particle excitations, as expected
for a Bose superfluid. Theoretical calculations have also shown
that, for droplets of this size, the discretized spectrum of 
collective excitations closely resembles the continuum of phonons
and rotons of bulk liquid \cite{casas,palma,rama,chin-MC}. 

Not all impurities are bound inside helium droplets; some prefer 
to stay at the surface. In this case, they can be sensitive to 
surface modes. For large droplets, theoretical calculations show 
that the spectrum of these excitations, with 
$l\ge 2$,  is close to the one of collective waves on a planar 
helium-vacuum interface (ripplons) \cite{chin-surface}. This 
also ensures that the liquid drop model (LDM) is accurate enough 
for many purposes. For example, using the liquid drop model for the 
excited states, together with a statistical method for the evaporation 
rate, the typical temperature of helium clusters in a supersonic beam 
was calculated in Ref.~\cite{brink}. The predicted values, for both 
$^3$He and $^4$He droplets, were later found to agree with 
experiments \cite{hartmann-T}, so providing an indirect insight
into the spectrum of surface excitations.  

The fact that in a uniform superfluid system single-particle excitations 
are not allowed to exist below the low-energy collective modes is
at the basis of the well known ``Landau criterion". This says that,
in order to create excitations in a superfluid by moving an impurity, 
one has to overcome some critical velocity, which is fixed by the 
spectrum of the excited states of the system. This follows from
the conservation of energy and momentum. In a uniform system the 
critical velocity is $v_c = {\rm min}[ \epsilon(p)/p ]$, 
where $\epsilon(p)$ is the energy of an excitations carrying 
momentum $p~$ \cite{Landau}. In liquid helium such a critical velocity 
has been the object of a longstanding investigation, involving the
structure of the phonon-roton branch and the nucleation and dynamics of 
quantized vortex lines and rings. Recently an experiment was realized 
to test the applicability of the Landau criterion to helium droplets 
\cite{harms-he3}. By studying low energy collisions of $^4$He and 
$^3$He atoms on $^4$He droplets, the authors found that $^3$He has
a significant probability to fly throughout the droplet without
dissipation of momentum and energy, and they interpreted this 
as a consequence of the Landau criterion.  The  existence of analog
critical velocities in dilute condensates is also currently 
investigated. In Ref.~\cite{MIT-Landau} an external potential (a 
laser beam) is used to produce a hole in a trapped Bose-Einstein  
condensate and this hole 
is moved back and forth at variable velocity, playing the role of 
a massive impurity. Theoretical interpretations in terms of the 
Landau criterion have been already presented in 
Refs.~\cite{adams,shlya}.

\subsection{Rotations and moment of inertia}

When a system is described by an order parameter of the form 
(\ref{eq:orderparameter}), a crucial consequence is the irrotationality
of the superfluid flow, whose velocity field is given by ${\bf v} = 
(\hbar / m)  \mbox{\boldmath$\nabla$} S$, the phase $S$ playing the 
role of velocity potential. This is a fundamental feature characterizing 
the superfluid motion, which applies to both liquid helium and dilute 
condensates. In this connection, it is worth recalling that, despite the 
fact that in a strongly correlated superfluid $|\Phi|^2$ differs from
the density $n$, the current associated with the superfluid 
velocity ${\bf v}$, at $T=0$, is still given by ${\bf j} = n {\bf v}$. 
The irrotationality of the superfluid flow has important manifestations
like, for instance, the peculiar behavior of the moment of 
inertia and the occurrence of quantized  vortices. 

The moment of inertia can be defined as the response, $ \Theta = 
\lim_{\Omega \to 0 } \langle L_z \rangle / \Omega $, to a rotating field 
of the form $H_{\rm rot}=-\Omega L_z$, where $L_z$ is the third component 
of the angular momentum operator. By rewriting the equations of hydrodynamics 
(\ref{eq:continuity})-(\ref{eq:euler}) in the frame rotating with 
angular velocity $\Omega$, one gets the irrotational moment of inertia 
\begin{equation}
\Theta =\left({ \langle x^2-y^2 \rangle \over \langle x^2+y^2 \rangle }
\right)^2 \Theta_{\rm rig}
\label {Thetairr}
\end{equation}
where $\Theta_{\rm rig}$ is the classical rigid value. This reduction of the 
moment of inertia compared to the rigid value was observed in liquid helium
in the 60's, so providing independent measurements of superfluid density 
\cite{hess}. In both helium droplets and trapped condensates, however, the
direct measurement of $L_z$ is not feasible and, consequently, the
determination of $\Theta$ requires more subtle probes. 

A suitable solution for trapped condensates is provided by the fact 
that, if the trap is deformed, the quadrupole and rotational degrees of 
freedom are coupled. This is well understood by taking a trap
with frequencies $\omega_x \ne \omega_y$ and considering 
the exact commutation relation $[H,L_z] = im(\omega^2_x-
\omega^2_y)Q$, which explicitly points out the link between the angular 
momentum operator $L_z$ and the quadrupole operator $Q=\sum_ix_iy_i$. 
Since the quadrupole variable can be easily excited and imaged in these
systems, one can use it to get information on $\Theta$. A successful
method consists in preparing a deformed condensate, whose shape at 
equilibrium is an ellipsoid with different major axis along $x$ and $y$, 
and suddenly rotating the magnetic trap by a certain angle around $z$.
The condensate then responds with an oscillation of its major axis 
in the $xy$-plane around the new equilibrium orientation. This is 
called {\it scissors} mode. Using the hydrodynamic equations for an
irrotational superfluid, one predicts the oscillation frequency $\omega 
= [\omega^2_x + \omega_y^2 ]^{1/2}$ \cite{David}, which is different from 
the result of a classical gas, $\omega = \mid \omega_x\pm\omega_y\mid$.
The explicit relationship between the scissors mode and the moment 
of inertia was derived in \cite{zamb}. Recent experiments \cite{Foot} 
have nicely confirmed these predictions. The observed moment of 
inertia clearly points out the superfluid nature of these 
condensates. 

In helium droplets, rotations can be induced and probed by acting on 
embedded impurities, namely, by studying their rotational transitions. 
Molecules like SF$_6$, OCS, HCN, and many others have been already used
(see Ref.~\cite{birgitta} for a recent review). Their spectrum inside 
the helium droplet exhibits sharp rotational lines, as those of free 
molecules, except for a different value of their moment of inertia. 
The sharpness of the spectrum is already a signature that the cold 
matrix around the molecule is superfluid. A nonsuperfluid matrix 
would smear the spectrum so much that it could not be resolved; a 
demonstration of this effect has been given in Ref.~\cite{grebenev} 
by replacing $^4$He (super) with a variable amount of $^3$He (normal).
In this case, the $^3$He component provides the low-lying single-particle 
(particle-hole) excitations which cause the broadening of the rotational 
spectrum of the molecule. On the other hand, the fact that the observed 
moment of inertia is systematically larger than the one of the free 
molecule (up to a factor 4 or 5 for some of them) is an interesting 
consequence of a dynamical coupling between the rotational degrees of 
freedom of the molecule and of the superfluid environment. Many 
experimental results are now available, but there are still open 
questions in their theoretical interpretation  
\cite{princeton,birgitta,kwon,lehmann}. One of the main
problems is that the coupling between the rotor and the superfluid
occurs on a microscopic lengthscale, comparable with both the
interatomic distances and the healing length. On this scale, one is
not fully ensured about the applicability of the superfluid 
hydrodynamic equations and/or of definitions of ``local" superfluid 
and normal densities.

\subsection{Rotations and quantized vortices}

Quantized vortices are striking manifestations of superfluidity. In 
liquid helium they have been the object of a longstanding investigation, 
starting from the pioneering ideas of Onsager and Feynman \cite{onsager} 
(see \cite{donnelly} for a detailed overview). 
As soon as BEC was observed in trapped gases,
the search for quantized vortices in these new systems became one of the 
primary goals. Recently, single vortices and arrays of vortices have
been obtained with different techniques: by ``phase imprinting" 
\cite{imprinting}, by using a laser beam ``stirrer" \cite{stirrer}, by 
rotating the magnetic trap \cite{rot-bucket}, and by rotating the
thermal cloud during the evaporative cooling process \cite{thermal}. 
Conversely, quantized vortices have not yet been observed in helium 
nanodroplets. 

The quantization of the circulation in a superfluid directly follows 
from the assumption that the order parameter (\ref{eq:orderparameter}) 
must be single-valued. Hence, the increment of the phase $S$ over any
closed path must be zero or an integral multiple of $2\pi$. If the 
superfluid  fills a simply-connected region, the irrotationality of the 
velocity field ($ \mbox{\boldmath$\nabla$} \times {\bf v} = 0$) strictly 
implies that the increment is zero, and the superfluid can not rotate
at all. In certain conditions, however, the fluid may prefer to develop 
vortical lines, where the superfluid density vanishes. The increment of 
$S$ around these lines can be an integral multiple of $2\pi$ and this 
implies that the circulation $\oint {\bf v} \cdot d{\bf l}$ is an 
integral multiple of the quantum $h/m$. 

The density distribution and the velocity field around a single 
rectilinear vortex line with circulation $h/m$ can easily be calculated 
in the case of dilute gases by using the Gross-Pitaevskii theory. In 
fact, a quantized vortex can be regarded as a stationary solution of 
Eq.~(\ref{eq:TDGP}) of the form
\begin{equation}
\Phi({\bf r},t)=\phi_v(r_\perp,z) e^{-i\mu t/\hbar } e^{i\varphi}  
\label{eq:phiv}
\end{equation}
where $\varphi$ is the azimuthal angle, $r_\perp$ is the distance from 
the $z$-axis, and $\psi_v$ is a real function obeying the equation
\begin{equation}
\left[ - {\frac{\hbar^2 \nabla^2 }{2m}} + {\frac{ \hbar^2 }{2m r_{\perp}^2}}
+ V_{\rm ext} (r_\perp,z) 
+ g \psi^2_v(r_\perp,z) \right] \phi_v(r_\perp,z) = \mu \phi_v(r_\perp,z) \;.
\label{eq:GPV}
\end{equation}
The velocity field associated with the order parameter (\ref{eq:phiv}) 
takes the form ${\bf v} = (\hbar/m)
\mbox{\boldmath$\nabla$}  \varphi = \hat{\bf z}\times 
{\bf r}/r^2$ where $\hat{\bf z}$ is the unit vector along $z$. 
It satisfies the irrotationality constraint everywhere except along the
vortical line and gives rise to  a total angular momentum given by 
$L_z = N\hbar$. The centrifugal term  $\propto 1/r^2_{\perp}$ in 
Eq.~(\ref{eq:GPV}) 
originates from the peculiar behaviour of the velocity field of the 
vortical configuration. This term is responsible for the vanishing of 
the condensate density $\phi_v^2$ along the $z$-axis. The ``core" 
of the vortex, i.e., the region where the density is depleted, has a
radius of the order of the healing length. The density profile and the
energy of the vortex can be calculated by numerically solving the
GP equation as done, for instance, by us in Ref.~\cite{ds}. The 
GP theory can also be used to investigate the nucleation mechanisms
and the stability of the vortex configurations (see \cite{fetter} for
a recent review); the theory describes the main features of the 
vortices observed in trapped BEC and several predictions agree well
with the available experimental results. 

What makes relatively easy to produce and manipulate vortices in 
trapped condensate is the possibility to induce rotations in a 
controllable way by means of laser beams or rotating traps. The
vortical lines can be directly observed as "holes" in the density
distribution. In fact, although the healing length is usually 
smaller than the resolution of the optical imaging devices, 
one can switch-off the confining potential, letting the condensate to
expand,  and wait till the expanding core is large enough to 
be observed \cite{vortexp}. 
Alternatively, one can use the lowest collective modes 
of the condensate in the trap as a probe of vorticity. For instance,
a vortical flow induces a frequency splitting of the two 
modes with $l=2$ and $m=\pm 2$, which results in a slow precession 
of the quadrupole shape deformations \cite{zambelli}, which is 
analog to the splitting of the lowest circularly polarized modes of 
the  vibrating wire in Vinen's ``milestone" experiment with  
superfluid helium \cite{vinen}.  This method has
been successfully used to estimate the angular momentum of 
vortical configuration \cite{chevy}, as well as to detect vortices
which are not directly observable because they are bent or tilted
\cite{haljan}. Finally, the $2\pi$ phase around a vortex 
can be also visualized in the form of a dislocation in the 
fringe pattern produced by the interference of two expanding  
condensates containing vortex lines \cite{fringes}. 

In helium droplets there is no external field to play with and
hence vortical configurations are not easy to produce and detect.
Vortices might nucleate in the initial stages of the jet 
expansion, where droplets form and cool down, or in the 
collisions with molecules in the pick-up scattering chamber; but 
no quantitative predictions have been given so far about these
processes. The stability of a vortex line in a droplet is also 
an open question. On the basis of purely energetic arguments, a 
vortex line is not stable since its formation costs a large amount 
of kinetic energy of the superfluid flow, and the ground state of the
droplets is hence always vortex-free \cite{Bau95}. On the one hand,
however, if a vortex appears as a metastable state in some
dynamical process, then one has to consider also the conservation
of both energy and angular momentum in any possible decay 
mechanism, and one might discover that this state is indeed 
long-living, despite its large energy, as it has been shown for 
vortex configurations in trapped condensates. On the other hand,
one can use dopant atoms or molecules to pin and stabilize the vortex, 
as the density functional calculations of Ref.~\cite{tn-ba} suggest.
In this case, the dopant can be used also as a probe of vorticity,
through the possible effects of the vortex flow on its rotational 
spectrum. An interesting perspective in this direction is the 
possibility to accommodate in the droplet a long chain of linear
molecules, such as HCN, as done in Ref.~\cite{nauta}. Such chains
are expected to be suitable for the stabilization of a vortex line
and, moreover, they can reveal in their spectrum the presence of
the vortex flow, like for a vibrating wire in a nanoscale Vinen-like 
experiment.

\section{Conclusions}
\label{sec:conclusions}

In this paper we reviewed some properties of both helium nanodroplets
and dilute trapped condensates. Our purpose was not to give a systematic 
and detailed account of the available theories and experiments, but 
rather to discuss some relevant features which allow one to build
a bridge between the two systems. We emphasized the existence of
a common theoretical background and of many analogies in the measured
properties, as well as of several interesting differences. We mainly 
used the language of Gross-Pitaevskii and density functional theory 
because, on the one hand, it represents a very reliable mean-field 
approach for dilute gases and, on the other hand, density functional 
approaches has been successfully used also in helium droplets. Of course, 
the close connections between trapped dilute gases and helium droplets
would equally emerge when  examined from the viewpoint of more 
{\it microscopic} theories or stochastic Monte Carlo techniques.  

Once this bridge is established, one may include most of the recent 
advances in both fields within a unified perspective, namely, the road
towards a deeper understanding of the physics of interacting quantum 
fluids.

\acknowledgements

F.D. thanks the Dipartimento di Fisica, Universit\`a di Trento
for the hospitality. This work is supported by MURST-COFIN2000.


\bigskip

\begin{figure}[t]
\epsfysize=8cm
\epsfxsize=10cm
\hspace{3cm}
\vspace{1cm}
\epsffile{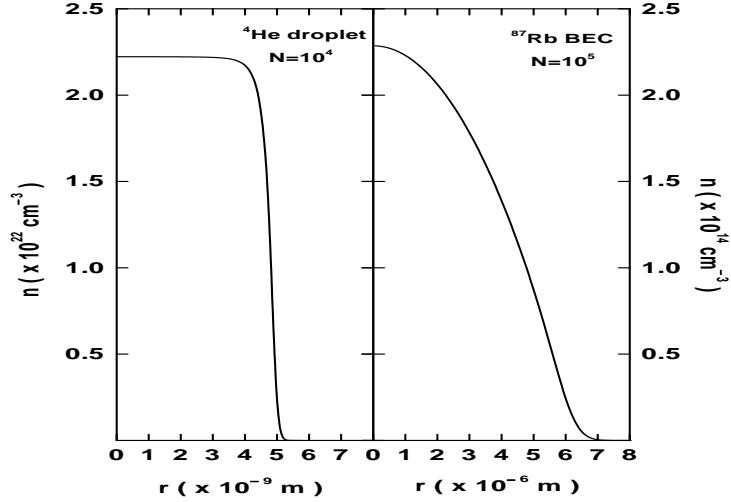}
\caption{ Left: density profile at $T=0$ of a $^4$He droplet obtained 
by solving Eq.~(\protect\ref{eq:hartree}) with the Orsay-Paris functional 
of Ref.~\protect\cite{Dup90}. Right: density
profile at $T=0$ of a condensate of $^{87}$Rb in a spherical trap with 
$a_{\rm ho}= 1.08 \times 10^{-6}$ m  obtained by solving the GP
equation (\protect\ref{eq:GP}).     }
\label{fig1}
\end{figure}

\begin{center}
\begin{tabular} {|c|c|c|c|c|}
\hline
\ &  
$a$  & 
$d$  & 
$\xi$  & 
$R$    \\
\hline 
$^4$He droplet  & & & &        \\  
$N=10^4$       & $\sim 3 \times 10^{-10}$ & $\sim 3 \times 10^{-10}$ 
& $\sim 10^{-10}$ & $ \sim 5\times 10^{-9}$    \\
\              & & & &        \\
\hline
$^{87}$Rb trapped BEC   & & & &        \\
$N=10^5$       & $\sim 5\times 10^{-9}$ 
& $\sim 10^{-7}$ & $\sim 10^{-7}$ & $\sim 5 \times 10^{-6}$    \\
$\omega_{\rm ho}= 2\pi (100 {\rm Hz}) $  & & & &        \\
$a_{\rm ho}= 10^{-6}$ m                  & & & &        \\
\hline
\end{tabular}
\end{center}
{\bf Table I}. Approximate lengthscales (in units of $m$) for a typical 
helium nanodroplet and a spherical trapped BEC. $a$: range of interaction 
(helium) and scattering length (trapped BEC). $d$: average interatomic 
distance. $\xi$: healing length, as in Eq.~(\protect\ref{eq:xi}). 
$R$: radius of the system. The corresponding 
density profiles are shown in Fig.~\protect\ref{fig1}. The energy per 
particle is about $6$ K for the helium droplet and $60$ nK for the 
condensate; the last 
value should be compared with  $\hbar  \omega_{\rm ho} \sim 5$ nK.

\begin{references}
\bibitem{noteT} We note on passing that one of the relevant topics 
about trapped gases is the kinetics of the formation of the condensate 
during the evaporation process. This might have significant analogies
with the formation of the superfluid phase (and the condensate) in 
helium droplets in the first stages of the evaporative cooling in 
the free jet expansion. From the conceptual viewpoint, this process 
is interesting and non-trivial, both for helium and trapped condensates.
\bibitem{note_anis} We only consider the case of isotropic
confinement, even though the traps used in current experiments can have
different trapping frequencies in $x,y$ and $z$ (most of them are
axisymmetric). The role of the anisotropy in the confinement is
important for specific purposes, but is not relevant for the present
discussion, where we privilege the comparison with spherical helium
droplets.
\bibitem{note-dipole}  Notice that the frequency of the dipole mode 
($n_r=0, \ell=1$) is the same for interacting and noninteracting 
particles, in agreement with the fact that this mode is a
center-of-mass oscillation of the whole condensate in the trap, 
having no effects in the internal degrees of freedom.
\end{references}
\end{document}